\documentclass{article}

\usepackage{arxiv}

\usepackage[utf8]{inputenc} % allow utf-8 input
\usepackage[T1]{fontenc}    % use 8-bit T1 fonts
\usepackage{hyperref}       % hyperlinks
\usepackage{url}            % simple URL typesetting
\usepackage{booktabs}       % professional-quality tables
\usepackage{amsfonts}       % blackboard math symbols
\usepackage{nicefrac}       % compact symbols for 1/2, etc.
\usepackage{microtype}      % microtypography
\usepackage{lipsum}
\usepackage{graphicx}
\usepackage{subfig}
\usepackage{amsmath}
\usepackage{algorithm}
\usepackage{algpseudocode}

\title{Evaluating Fitness Averaging Strategies in Cooperative NeuroCoEvolution for Automated Soft Actuator Design}

\author{
    Hugo Alcaraz-Herrera\\
    Unconventional Computing Laboratory, \\
    College of Arts, Technology and Environment, \\
    University of the West of England, \\
    Bristol, BS16 1QY, United Kingdom \\ \texttt{hugo.alcaraz@uwe.ac.uk}
\And
    Michail-Antisthenis Tsompanas  \\
    Unconventional Computing Laboratory \& \\
    School of Computing \& Creative Technologies,\\
    College of Arts, Technology and Environment, \\
    University of the West of England,\\ 
    Bristol, BS16 1QY, United Kingdom \\
    \texttt{antisthenis.tsompanas@uwe.ac.uk}
\And
       Igor Balaz\\
       Laboratory for Meteorology, Physics and Biophysics,\\ Faculty of Agriculture, \\
       University of Novi Sad, \\ Trg Dositeja Obradovica 8, 21000, Novi Sad, Serbia
\And
       Andrew Adamatzky \\
       Unconventional Computing Laboratory,\\
    College of Arts, Technology and Environment, \\
    University of the West of England,\\ 
    Bristol, BS16 1QY, United Kingdom \\
}

\begin{document}
\maketitle

\begin{abstract}
Soft robotics are increasingly favoured in specific applications such as healthcare, due to their adaptability, which stems from the non-linear properties of their building materials. However, these properties also pose significant challenges in designing the morphologies and controllers of soft robots. The relatively short history of this field has not yet produced sufficient knowledge to consistently derive optimal solutions. Consequently, an automated process for the design of soft robot morphologies can be extremely helpful. This study focusses on the cooperative NeuroCoEvolution of networks that are indirect representations of soft robot actuators. Both the morphologies and controllers represented by Compositional Pattern Producing Networks are evolved using the well-established method NeuroEvolution of Augmented Topologies (CPPN-NEAT). The CoEvolution of controllers and morphologies is implemented using the top $n$ individuals from the cooperating population, with various averaging methods tested to determine the fitness of the evaluated individuals. The test-case application for this research is the optimisation of a soft actuator for a drug delivery system. The primary metric used is the maximum displacement of one end of the actuator in a specified direction. Additionally, the robustness of the evolved morphologies is assessed against a range of randomly generated controllers to simulate potential noise in real-world applications. The results of this investigation indicate that CPPN-NEAT produces superior morphologies compared to previously published results from multi-objective optimisation, with reduced computational effort and time. Moreover, the best configuration is found to be CoEvolution with the two best individuals from the cooperative population and the averaging of their fitness using the weighted mean method.
\end{abstract}

% keywords can be removed
\keywords{Soft Robots  \and NeuroCoEvolution \and NeuroEvolution of Augmented Topologies \and Compositional Pattern Producing Networks}

\section{Introduction}\label{sec:introduction}

Soft robotics is a branch of robotics that focuses on machines whose building materials are flexible and ductile (e.g., silicone and gel). Due to their morphology and movements being mainly inspired by living organisms in Nature, soft robots have shown better performance than traditional robots in specific tasks, for instance, medicine and healthcare \cite{Hsiao2019}. Nonetheless, to design the physical structure of a novel soft robot entails significant challenges, since the materials that build them are characterised by non-linear mechanical properties that are difficult to model \cite{Hiller2014}. Moreover, building a significant amount of prototypes to test and optimise the design of these machines is not a viable solution, because of the time and material resources required for such a task \cite{Schulz2016}.

%\textcolor{red}{Building soft robots is an intricate task because their materials present non-linear mechanical properties that are complex to characterise \cite{Hiller2014}. Consequently, finding an adequate design (i.e., morphology) implies considerable time and material resources since numerous prototype models need to be tested in ``the real world'' \cite{Schulz2016}. }

Another key step in the pipeline of producing an efficient soft robot is determining the controller that will drive it. This usually happens after the physical structure is finalised. This step is also significantly complicated for soft robots due to their flexibility that hinders precise control and feedback mechanisms \cite{Wang2022}. Consequently, an automated process for designing the physical structure and the controller strategy would enable the acceleration of designing novel soft robots. Cooperative NeuroCoEvolution of soft robot actuators is proposed here and the fitness of each morphology and controller pair is provided by a physics simulator. The simulator used is Voxelyze \cite{Hiller2014}. The representation of possible morphologies and controllers is based on an indirect method of using  Compositional Pattern Producing Networks (CPPNs) \cite{Stanley2007cppn}.

%\textcolor{red}{Once an adequate soft robot model is found, the next step is designing a suitable controller for it, which is typically as complex as the process of finding a suitable morphology design. This is due to the fact that the materials used can deform at every point, which implies infinite degrees of freedom, including contraction, torsion, extension, and bending. Furthermore, they present time-dependent properties. These features complicate the process of modelling the behaviour and movement of soft robots \cite{Wang2022}.}

A well-established method for training neural networks is NeuroEvolution (NE), which  applies the concept of genetic algorithms on the topology and weights of neural networks. The most prominent example of NE methods is the NeuroEvolution of Augmenting Topologies (NEAT) \cite{Stanley2002}, which was successfully utilised to separately sketch the physical structures of robots \cite{Auerbach2011} and their controllers \cite{Alcaraz2024controllers}.

%%CPPN-NEAT?
%\textcolor{red}{A promising strategy to address the inherent design challenges of soft robots is neuroevolution (NE), which evolves the topology of artificial neural networks (ANNs) through a genetic algorithm (GA). One of the most popular NE-based approaches is Neuroevolution of Augmenting Topologies (NEAT) \cite{Stanley2002}, which has been implemented to generate robot morphologies \cite{Auerbach2011}, and control design \cite{Alcaraz2024controllers}.}

CoEvolution has been shown to provide fitter individuals for diversified targets compared to traditional evolutionary optimisation, i.e., feature weighting \cite{Blansch2005}. Thus, the amalgamation of these techniques (termed here as NeuroCoEvolution) is investigated in terms of using different measures of central tendency to assign a fitness value for one individual of the evolving population cooperating with multiple individuals of the other population. In specific, different definitions of mean values were tested, such as arithmetic, weighted, geometric and harmonic mean. Moreover, several scenarios for the cooperation scheme between the evolving individual and the $n$ fittest individuals from the static population were investigated. The considered test-case involves optimising soft actuator morphologies (SAMs) and their controllers under the assumption of using them in medical devices that would facilitate targeted drug delivery to anatomically inaccessible regions of the human body.

Thus, the primary objectives of this research are: (i) analysing the suitability of NEAT to evolve CPPNs under a cooperative CoEvolutionary scheme \cite{Ma2019} for designing SAMs and their controllers, (ii) identifying the optimal cooperative configuration through the exploration of four evaluation schemes based on the arithmetic mean, weighted mean, geometric mean, and harmonic mean, respectively; and (iii) comparing the performance of the NeuroCoEvolutionary approach against a NeuroEvolutionary and a multi-objective-based approaches.

%\textcolor{red}{Since there is evidence that coevolution can perform better than evolution in diverse tasks such as feature weighting \cite{Blansch2005}, the primary objective of this research is assessing the suitability of NEAT under a cooperative coevolutionary scheme \cite{Ma2019} for designing soft actuator morphologies (SAMs) and their controllers. SAMs can be built using biological material (e.g., cells or tissue) with a purpose of delivering drugs in areas of the human body that are difficult to reach. The analysis here focuses on the capability of SAMs to make upward bending movements induced by the controllers being evolved within the same methodology.}

%The rest of the paper is organised as follows: Section~\ref{sec:background} describes research focused on soft robot related tasks employing NEAT. Furthermore, research related to coevolution is also depicted. Section~\ref{sec:algorithms} introduces in NEAT and Age-Fitness Pareto Optimisation (AFPO) a popular multi-objective optimisation algorithm, which is used as a baseline for experimentation. Section~\ref{sec:setup} presents the relevant aspects of the experimental phase described in this research. Section~\ref{sec:experimental} describes in detail the experiments performed in the task of designing SAMs and their controllers. Finally, Section~\ref{sec:conclusions} concludes this work, presents insights from the research, and outlines future work.

%%%%%%%%%%%%%%
% BACKGROUND %
%%%%%%%%%%%%%%

%=======
\section{Background}\label{sec:background}

NEAT has been utilised in prior work as a generative mechanism for designing soft robot morphologies. In \cite{Auerbach2010}, NEAT is used to generate three-dimensional structures that can conserve momentum to reach maximum displacement due to gravity. The structures are composed of spherical cells that fuse to form rigid bodies. The growth process begins with a root, namely, a single cell positioned at the origin of the structure, around which a cloud of $n$ points is initialised. The next step is querying the ANN, whose output determines where to allocate a cell. Results point out that the structures created by this method can embody the intricate relationship between physical structure and functional behaviour, which is not readily apparent.

Another example can be found in \cite{Cheney2015}, where soft robots capable of passing or squeezing through a small aperture are designed using NEAT. The experiments focused on simulations where each robot starts within a cage made of voxels (the elementary building block in the three-dimensional simulator used). The cage dimensions are 15 in the $x$ and $y$ axes and 11 voxels in the $z$ axis, leaving a gap of one voxel in the $x$ and $y$ axes between the edge of the cage. The maximum size of robots is 11 on each axis. The cage is static and rigid. Each side has a circular opening of diameter 10 voxel lengths. The results demonstrated that robots with greater deformability exhibit superior capability in navigating through confined spaces, such as narrow tunnels, compared to their more rigid counterparts.

Furthermore, NEAT can discover soft robot morphologies capable of performing more than one task \cite{Kimura2021}. The research argues that by mixing the genotype of soft robots focused on a single task, it is possible to create multi-functional soft robots. The experimental methodology employed in this study involved simulations of voxel-based creatures performing tasks in both aquatic and terrestrial environments. The results indicate that the proposed method effectively explores the morphological search space and is capable of simultaneously satisfying two tasks more efficiently than existing approaches.

On the other hand, NEAT has also been used to design controllers for soft robots. For instance, in \cite{Wen2017}, a controller for a three-link planar arm model driven by nine muscles is developed. The training sets include the geometry relationships, forward kinematics, and muscle mechanic equations. A comparison between NEAT and a fixed-topology ANN revealed that the NEAT-evolved controller achieves superior performance in targeted tasks, including precise arm positioning.

Moreover, NEAT was proposed to control robots in crowded environments \cite{Seriani2021}. That research states that a ray-casting model can be employed as the perception mechanism of robots for industrial purposes. The ray-casting model can detect objects at the current time and has a memory of previously sensed objects. The experiments related to this work focus on assessing simulated versions of the controllers and then physically, in a minimal implementation. Results advocate that controllers designed by NEAT exhibit a suitable performance in both evaluation scenarios.

In addition, in \cite{Caceres2017}, an implementation of NEAT is introduced to control autonomous vehicles. A reactive navigation hybrid controller for a non-holonomic mobile robot is used for experimentation in a simulation platform developed by the authors. The simulator considers the kinematic model of a vehicular robot. Results indicate that controllers generated by NEAT show a suitable performance in unknown environments, reaching the target point and avoiding obstacles. 

Cooperative CoEvolution has been implemented to optimise the behaviour of real robots in the pursuit domain \cite{Sun2020}. In each population, the first individual represents a real robot, while the remaining individuals are virtual robots that function as an action space, enabling the real robot to explore and sample its local environment. The experimental methodology evaluates the generality, stability, and scalability of the proposed approach across four distinct prey movement scenarios: stationary prey, linearly moving prey, randomly moving prey, and linear smart prey. The results indicate that the proposed approach consistently achieves successful prey capture without requiring additional modifications.

Another example of implementing cooperative CoEvolution is described in \cite{Guettas2014}. The research is oriented to design configurations and controllers for robots that implicitly support self-configuration. The algorithm introduces hints to facilitate the search for suitable solutions. One population is composed of sets of sequences of movements that can rearrange a specific modular configuration into a new one; the other population contains ANNs with fixed topology representing controllers to perform locomotion tasks. Experiments are conducted in a simulated environment implemented with a physics engine called {\em PhysX}. Results point out that coevolving both robot configurations and controllers enhances the performance of robots and optimises their locomotion behaviour.

Furthermore, cooperative CoEvolution has been used to optimise the policy parameters of agents individually while they are interacting as a team \cite{Fontbonne2022}. The proposed algorithm operates on team composition by constantly updating a limited number of agents, considering the complexity of the current task. Another feature of this algorithm is the modulation of the number of new policies added to the current team through time, which provides efficient learning regardless of the time without requiring knowledge of the problem {\em a priori}. Experiments were conducted using the well-known El Farol Bar problem, commonly employed in collective decision-making research. The problem was adapted to represent a multi-agent resource selection scenario, in which the approach demonstrated satisfactory performance.

Therefore, considering the previously described research findings, it is feasible to consider NEAT as a solid design engine for SAMs and their controllers under a cooperative CoEvolutionary scheme to enhance the search in both search spaces. 

%%%%%%%%%
% NEAT %
%%%%%%%%

%=======
\section{NeuroEvolution of Augmented Topologies}\label{sec:neat}

Arguably, NEAT \cite{Stanley2002} is one of the most successful NE algorithms. It was developed to address three key limitations identified in earlier algorithms: (a) the absence of solution representations that support the recombination of arbitrary network topologies; and (b) the tendency for novel features (i.e., newly introduced elements in network topologies) to disappear early in the evolutionary process; and (c) preventing punishing complex network topologies by using appropriate fitness functions. Thus, NEAT has three main features:

\begin{enumerate}
\item {\em Gene tracking} using a list of connection genes representing a connection between two nodes that contains the ``origin'' and ``destination'' node, the weight of the connection, the state of the connection (enabled or not), and a unique numeral ID called {\em global innovation number}.
\item {\em Speciation to protect innovation} which creates species within population whose purpose is to induce interactions among individuals of the same species and limiting species to expand over the population by utilising explicit fitness sharing.
\item {\em Minimising network structures} by initialising individuals connecting in full the input neurons to the output neurons. In this way, new neurons are incrementally added to generate new topologies
\end{enumerate}

\begin{figure}[tb!]
  \centering
     \includegraphics[width=1.0\linewidth]{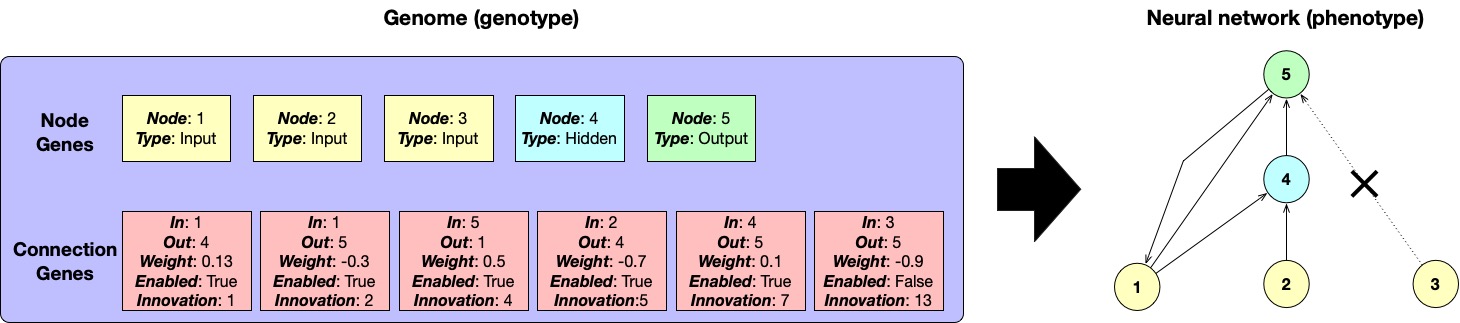}
  \caption{Example of a genotype and its phenotype under NEAT. Figure adopted from \cite{Alcaraz2024actuator}.}
  \label{fig:neat_gen_phe}
\end{figure}

These features are meant to counteract the issues observed in previous approaches. Figure~\ref{fig:neat_gen_phe} depicts an example of a genotype and phenotype of an ANN under the NEAT representation strategy. In general, NEAT can be applied to different types of neural networks. For instance the evolution of CPPNs, which is the method implemented here, is termed CPPN-NEAT.

%%%%%%%%%%%%%%%%%%%%%%
% EXPERIMENTAL SETUP %
%%%%%%%%%%%%%%%%%%%%%%

%=======
\section{Experimental setup}\label{sec:setup}

Three key elements are essential to conduct the experiments outlined in the following: (i) an environment where SAMs are simulated; (ii) the specific configuration employed for CPPN-NEAT during experimentation; and (iii) the NeuroCoEvolution setup implemented. To enhance transparency and enable reproducibility, the following sections outline key aspects of the implementation.

%These key aspects needed to build the experimental setup of this research are described in the following sections.

%=====
\subsection{Voxelyze}\label{sec:setup_voxelyze}

This research aims to assess the suitability of CPPN-NEAT in a cooperative NeuroCoEvolutionary approach for the joint design of SAMs and their controllers. These devices can play the role of actuators on in steerable surgical catheters. Due to the dynamics of SAMs, a three-dimensional space ($x,y,z$) is used as a design canvas. Furthermore, one end of the SAMs is fixed (i.e., unable to move in any case) to represent the insertion point of the catheter in this given example.

The {\em Voxelyze} simulator \cite{Hiller2014,KriegmanGitHub} is an open-source virtual environment utilised to predict the mechanical behaviour of SAMs considering physical aspects of the environment such as viscosity and gravity. The fundamental unit of this platform is the \textit{voxel}, which can be configured to represent various material properties. Thus, a SAM is composed of stacking voxels. In this study, two distinct materials are represented by specific voxel types: active and passive, each corresponding to the response of contractile and immotile building blocks. Following previous {\em in vitro} works \cite{Kriegman2020}, the parameters set for Voxelyze are: Young's modulus to $5\times10^6$, Poisson ratio to $0.35$, coefficients of static and dynamic friction set to $1.0$ and $0.5$, respectively. Regarding active voxels, they were characterised by volumetric actuation of $\pm 50\%$ of their at rest volume with a frequency of $4$ Hz.

Voxelyze is used as the core element to provide the fitness function of SAMs and their controllers. The original open-source code was modified for the physics engine to trace the position of the free end of morphologies in the ($x,y,z$) during a time $t$. %The free end only has one degree of freedom in the $yz$ plane (i.e., only a vertical movement). 
Following the findings of previous research \cite{Tsompanas2024}, SAMs have a passive enclosure used to represent a bioreactor required for a muscle actuator with a nutrient supply. The output of this modified version of Voxelyze consists of the trace of the free end in the three-dimensional space ($x,y,z$) of the SAM being simulated from the starting position ($t=0$) and the final position ($t=n$).

Finally, the dimensions of the SAMs, expressed in voxel units, are 20 along the $x$-axis and 8 along both the $y$- and $z$-axes. Consequently, the three-dimensional canvas spans the range [0,20] along the $x$-axis, and [0,8] along the $y$- and $z$-axes.

%=====
\subsection{CPPN-NEAT configuration}\label{sec:setup_neat}

An extended version of NEAT called {\em CPPN-NEAT} \cite{Stanley2007cppn} is utilised in this research. CPPN-NEAT evolves a specific type of ANN called CPPN \cite{Stanley2007cppn}. One of the primary capabilities of CPPNs is that they have a wider set of functions in hidden neurons, such as sine, cosine, and quadratic, not only sigmoid or RELU, as in standard ANNs. This capacity allows CPPNs to generate diversified patterns (i.e., symmetry) in their outputs.  

This research uses a two-population cooperative NeuroCoEvolution system. One population is composed of CPPNs encoding SAMs, whereas the other contains CPPNs encoding controllers. % Therefore, CPPN-NEAT was configured in two ways: one for SAMs and the other for controllers. 
Both configurations are described as follows: 

%===
\subsubsection{SAM configuration.}\label{sec:setup_neat_sam}

Voxelyze, the physics engine used to simulate SAMs (see Section~\ref{sec:setup_voxelyze}), operates in a discrete three-dimensional canvas. Consequently, for each point $i$ on the canvas, it is necessary to specify both the presence or absence of a voxel ($\nu_i$) and its material type ($m_i$). Based on this, CPPNs encoding SAMs are queried as follows:

\begin{equation}\label{eq:setup_neat_sam} 
    CPPN_{sam}(x_i,y_i,z_i) = \nu_i,m_i
\end{equation}

\noindent
where the $(x_i, y_i, z_i)$ tuple is the coordinates of the $i$-th point in the canvas, $v_i$ refers to the presence (or not) of a voxel at the $i$-th point in the canvas, and $m_i$ represents the material type of the voxel at the $i$-th point of the canvas. As previously mentioned (see Section~\ref{sec:setup_voxelyze}), two types of materials are employed during experimentation: {\em passive}, which is encoded as $P$, and {\em contractile}, encoded as $C$. Regarding $\nu_i$, it is processed through the following equation:

\begin{equation}\label{eq:setup_neat_sam_v}
    Presence(x_i,y_i,z_i) = \begin{cases}
    \text{yes,  } & |v_i| \geq 0.5 \\ 
    \text{no, } & \text{otherwise}
    \end{cases}
\end{equation}

On the other hand, $m_i$ is processed as follows:

\begin{equation}\label{eq:setup_neat_sam_m}
    Material~code(x_i,y_i,z_i) = \begin{cases} 
    P\text{,  } & |m_i| < 0.5 \\ 
    C\text{,  } & \text{otherwise}              
    \end{cases}
\end{equation}

%===
\subsubsection{Controller configuration}\label{sec:setup_neat_controller}

Considering that SAMs are simulated in a discrete three-dimensional canvas, and the information about their type of material is available, the input of CPPNs is composed by the coordinates of the $i$ point and the material type of the voxel situated at the $i$ point of the canvas. The output of CPPNs contains the phase offset of the contraction of active voxels situated at the $i$ point in the canvas. In this context, the phase offset denotes the delay in the expansion behaviour of active voxels. Consequently, CPPNs encoding the controllers are queried as follows:

\begin{equation}\label{eq:setup_neat_controller} 
    CPPN_{con}(x_i,y_i,z_i,m_i) = pho_i
\end{equation}

\noindent
where the $(x_i, y_i, z_i)$ tuple represents the coordinates of the $i$-th point of the three-dimensional canvas, $m_i$ is the code of the material type of the voxel allocated at the $i$-th of the canvas (see Equation~\ref{eq:setup_neat_sam_m}). The variable $pho_i$ is the phase offset of the voxel at the $i$-th point of the canvas. According to the findings of previous research \cite{Alcaraz2024controllers}, the output of CPPNs is clamped to $[-2\pi,2\pi]$ to allow complete sinusoidal-like contraction movements.

%=====
\subsection{CoEvolutionary setup}\label{sec:setup_coevolutionary_setup}

The populations composing the CoEvolutionary system (i.e., SAMs and controllers) have 25 individuals each. Furthermore, for all approaches, each evolutionary run lasted 200 generations. For each one generation a new set of 25 individuals are produced for one of the two populations. 
Each CPPN (i.e., individual) is initialised with a minimal topology which comprises no hidden neurons and direct, fully connected pathways from input to output neurons.

The activation function dictionary utilised for experimentation contains: {\em sine}, {\em negative sine}, {\em absolute value}, {\em negative absolute value}, {\em square}, {\em negative square}, {\em squared absolute value}, {\em negative squared absolute value}, {\em sigmoid}, {\em clamped}, {\em cubical}, {\em exponential}, {\em Gaussian}, {\em hat}, {\em identity}, {\em inverse}, {\em logarithmic}, {\em ReLU}, {\em SeLU}, {\em LeLU}, {\em eLU}, {\em softplus}, and {\em hyperbolic tangent}. The evolutionary parameters used for CPPN-NEAT are shown in Table~\ref{tab:setup_coevolutionary_setup}. It is essential to note that these parameters were taken directly from \cite{Alcaraz2024actuator} and not modified, as this facilitates a meaningful comparison during experimentation (see Section~\ref{sec:experimental_robustness}). Therefore, no fine calibration was performed, as the scope of this research is focused on finding the optimal cooperative configuration. 

\begin{table}
\caption{Parameters utilised for CPPN-NEAT during experimentation.}
\label{tab:setup_coevolutionary_setup}
 \begin{center}
  \begin{tabular}{  |c|c|p{8cm}| } 
\hline
   Parameter & Value & Description \\
\hline
   Compatibility threshold & 3 & Individuals whose {\em genomic distance} is less than this value belong to the same species \cite{Stanley2002}.\\
   Compatibility disjoint coefficient & 1.0 & Contribution coefficient for the disjoint and excess count of genes during the calculation of the genomic distance of individuals.\\ 
   Compatibility weight coefficient & 0.5 & Contribution coefficient for each bias and weight during the calculation of the genomic distance of individuals.\\
   Maximum stagnation & 25 & A species must present improvement in not more than this number of generations. Otherwise, it is removed.\\
   Survival threshold & 0.6 & The proportion of each species to reproduce during evolution. \\ 
   Activation function mutate rate & 0.4 & Replacing the activation function of neurons occurs with this probability.\\
   Add/delete connection rate & 0.3/0.2 & Adding/deleting a connection between existing neurons takes place with these probabilities.\\
   Activate/deactivate connection rate & 0.5 & Activating/Deactivating an existing connection between neurons occurs with this probability.\\
   Add/delete node rate & 0.3/0.2 & The probabilities to add/delete a neuron.\\
\hline
  \end{tabular}
 \end{center}
\end{table}

Furthermore, the cooperative NeuroCoEvolution scheme between populations works in a round-robin fashion: the same computational resources are assigned to each population \cite{Ma2019}. In addition, based on the insights of previous research \cite{Alcaraz2025coevolution}, the collaboration strategy implemented is {\em n fittest individuals vs all}, which consists of employing the $n$ best individuals of one population to evaluate all the individuals of the other population. Moreover, the set of values used for $n$ during experimentation is $\{2, 3, 5, 7, 10\}$.

Four different evaluation approaches (i.e., fitness functions) of the central tendency for the collaborators are implemented:

\subsubsection {\em Arithmetic mean} (AM). Individuals are evaluated using the following equation:

\begin{equation}\label{eq:setup_coevolutionary_arithmetic} 
    apt_a = \frac{\displaystyle\sum_{i=1}^{n} \delta_i}{n}
\end{equation}

\noindent
where $apt_a$ is the aptitude of the individual $a$ in the evaluated population, whereas $\delta_i$ represents the displacement observed in the $yz$ plane of the SAM $i$ being simulated. Furthermore, $n$ is the number of fittest individuals taken from the population not being evaluated

\subsubsection {\em Weighted mean} (WM). The aptitude ($apt$) of an individual $a$ is calculated by first sorting in descending order (i.e., the largest number at the beginning of the list and the smallest at the end) all the displacements observed in the $yz$ plane ($\delta$) of the morphology $i$ being simulated. Then, the following equation is applied:

\begin{equation}\label{eq:setup_coevolutionary_weighted} 
    apt_a = \sum_{i=1}^{n} \delta_i \times \omega_i
\end{equation}

\noindent
where $\omega_i$ is the $i$-th element of the weight vector. Different weights are implemented for each value of {\em n}. Table~\ref{tab:setup_coevolutionary_weighted} presents the values used to calculate the weighted mean in different approaches. As a an example, when $n=3$, the aptitude of an individual $m$ is calculated as: $apt_m = \delta_1(0.5) + \delta_2(0.3)+ \delta_3(0.2)$. 
    
\subsubsection {\em Geometric mean} (GM). Individuals are evaluated through the following equation:

\begin{equation}\label{eq:setup_coevolutionary_geometric} 
    apt_a = \sqrt[n]{\prod_{i=1}^{n} \delta_i}
\end{equation}

\noindent
where $apt_a$ is the aptitude of the individual $a$ in the evaluated population, whereas $\delta_i$ represents the displacement observed in the $yz$ plane of the SAM $i$ being simulated. Furthermore, $n$ is the number of fittest individuals taken from the population not being evaluated.

\subsubsection {\em Harmonic mean} (HM). The aptitude of an individual $a$ ($apt_a$) is calculated by employing the following equation:

\begin{equation}\label{eq:setup_coevolutionary_harmonic} 
    apt_a = \frac{n}{\displaystyle\sum_{i=1}^{n} \frac{1}{\delta_i}}
\end{equation}

\noindent
where $\delta_i$ is the displacement observed in the $yz$ plane of the SAM $i$ being simulated. Furthermore, $n$ represents the number of fittest individuals taken from the population not being evaluated.

It is important to highlight that these evaluation approaches were selected since they have different statistical features that affect the calculation of the aptitude of individuals during evolution. For instance, GM and HM are not significantly affected by outlier values, whereas AM is significantly affected \cite{manikandan2011measures}. 

\begin{table}
\caption{Weights used to evaluate individuals employing different values for {\em n} under WM.}
\label{tab:setup_coevolutionary_weighted}
 \begin{center}
  \begin{tabular}{ c c } 
\hline
   {\em n} & Weights ($\omega$) \\
\hline
   2 & [0.6, 0.4] \\
   3 & [0.5, 0.3, 0.2] \\
   5 & [0.4, 0.3, 0.15, 0.1, 0.05] \\
   7 & [0.35, 0.25, 0.15, 0.12, 0.07, 0.04, 0.02] \\
   10 & [0.3, 0.2, 0.15, 0.12, 0.08, 0.05, 0.04, 0.03, 0.02, 0.01] \\ 
\hline
  \end{tabular}
 \end{center}
\end{table}

\subsection{Infrastructure}\label{sec:setup_infrastructure}

Since simulations of SAMs imply a significant computational time, the experimental infrastructure implemented in \cite{Alcaraz2024locomotion,Alcaraz2024actuator} was replicated. Thus, the hardware configuration is described as follows: {\em Processor}: ARM (virtualised), nine cores (18 threads), 3.20 GHz. {\em RAM Memory}: 16 GB, LPDDR5, where CPPN-NEAT was implemented using a client–server architecture to leverage the advantages of distributed computation.

%%%%%%%%%%%%%%%%%%%%%%%%
% EXPERIMENTAL RESULTS %
%%%%%%%%%%%%%%%%%%%%%%%%

%=======
\section{Experimental results}\label{sec:experimental}

The performance of the cooperative NeuroCoEvolution system is assessed through two experiments: (i) determining the optimal value of $n$ and the evaluation approach (i.e., the optimal cooperative configuration) in terms of finding the SAM with the maximum displacement possible in the $yz$ plane, and (ii) studying the robustness of the fittest SAMs discovered by the optimal cooperative configuration discovered in the previous experiment employing numerous controllers. For each cooperative configuration, 30 evolutionary trials were conducted.

%=====
\subsection{Optimal cooperative configuration}\label{sec:experimental_coop_conf}

\begin{figure}[tb!]
  \centering
     \includegraphics[width=1.0\linewidth]{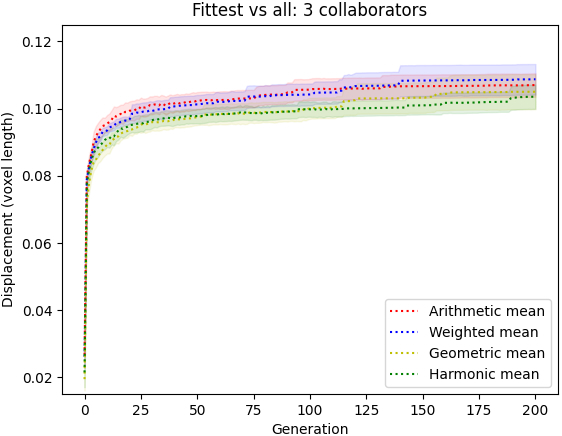}
  \caption{Mean performance observed of the fittest SAM using AM, WM, GM, and HM showing 95\% confidence interval (shaded region) utilising 3 collaborators.}
  \label{fig:experimental_coop_conf_3}
\end{figure}

\begin{figure}[tb!]
  \centering
     \includegraphics[width=1.0\linewidth]{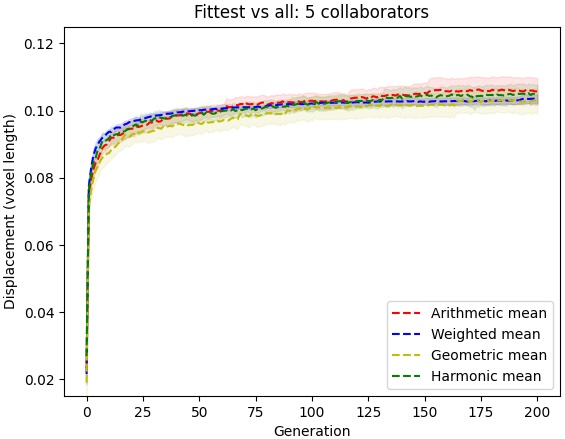}
  \caption{Mean performance observed of the fittest SAM using AM, WM, GM, and HM showing 95\% confidence interval (shaded region) utilising 5 collaborators.}
  \label{fig:experimental_coop_conf_5}
\end{figure}

\begin{figure}[tb!]
  \centering
     \includegraphics[width=1.0\linewidth]{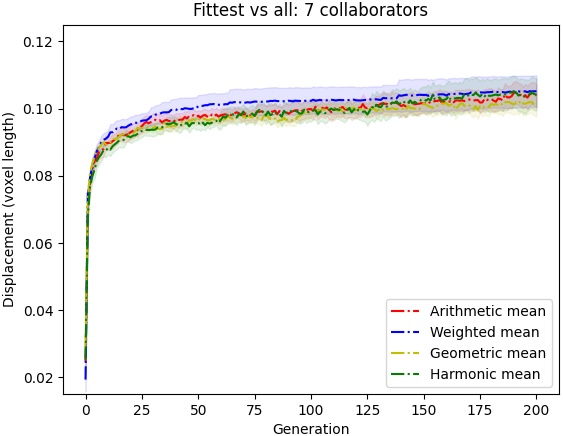}
  \caption{Mean performance observed of the fittest SAM using AM, WM, GM, and HM showing 95\% confidence interval (shaded region) utilising 7 collaborators.}
  \label{fig:experimental_coop_conf_7}
\end{figure}

\begin{figure}[tb!]
  \centering
     \includegraphics[width=1.0\linewidth]{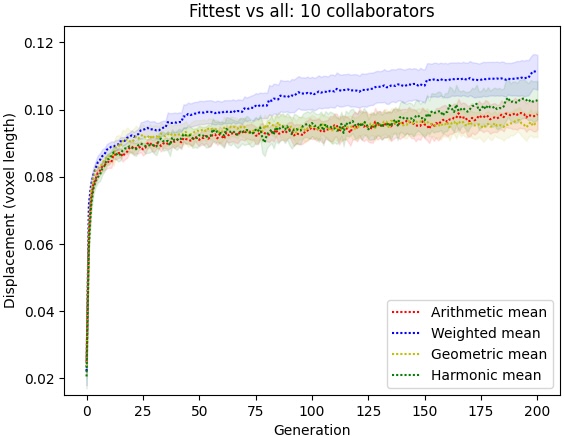}
  \caption{Mean performance observed of the fittest SAM using AM, WM, GM, and HM showing 95\% confidence interval (shaded region) utilising 10 collaborators.}
  \label{fig:experimental_coop_conf_10}
\end{figure}

This experiment aims to identify the cooperative configuration capable of discovering SAMs (and their controllers) that can exhibit a significant displacement in the $yz$ plane. Figures~\ref{fig:experimental_coop_conf_3} to \ref{fig:experimental_coop_conf_10} illustrate the mean performance of the fittest individual (i.e., the fittest SAM), with shaded regions indicating 95\% confidence intervals, based on 30 evolutionary runs under AM, WM, GM, and HM, and the values considered for $n$ (see Section~\ref{sec:setup_coevolutionary_setup}).

%\begin{figure}[tb!]
%  \centering
%     \includegraphics[width=1.0\linewidth]{experimental_coop_conf.jpg}
%  \caption{Mean performance observed of the fittest SAM using AM, WM, GM, and HM showing 95\% confidence interval (shaded region) utilising: (a) 2 collaborators; (b) 3 collaborators; (c) 5 collaborators; (d) 7 collaborators; and (e) 10 collaborators.}
%  \label{fig:experimental_coop_conf}
%\end{figure}

\begin{figure}[tb!]
  \centering
     \includegraphics[width=1.0\linewidth]{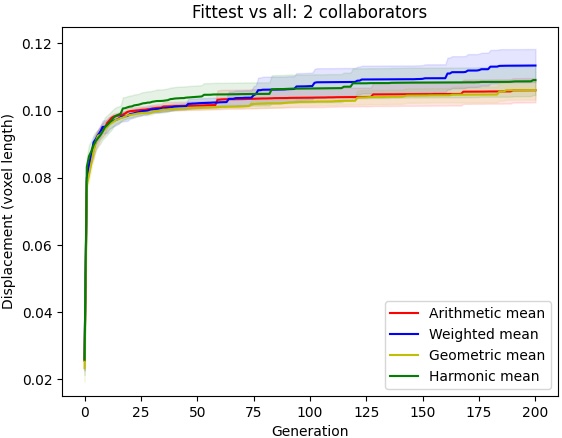}
  \caption{Mean performance observed of the fittest SAM using AM, WM, GM, and HM showing 95\% confidence interval (shaded region) utilising 2 collaborators.}
  \label{fig:experimental_coop_conf_2}
\end{figure}

When $n=2$ (see Figure~\ref{fig:experimental_coop_conf_2}), all the evaluation approaches exhibit a clear tendency to evolve, each with a different pace. Furthermore, analysis revealed statistically significant performance differences. The Shapiro–Wilk test confirmed that the data did not follow a normal distribution ($p < 0.05$). Then, by employing Dunn's test, it is feasible to confirm that no significant differences between AM and GM exist ($p>0.05$). In contrast, there are significant differences among WM and HM. Hence, it is possible to rank the performance of the four evaluation approaches: WM $>$ HM $>$ AM and GM.

Furthermore, when $n=3$ (see Figure~\ref{fig:experimental_coop_conf_3}), again, the evaluation approaches present an evident evolutionary behaviour. However, in the last generations, the evolutionary process stagnates. In addition, some significant differences are observed among the evaluation approaches' performances. All the data collected are not normally distributed (Shapiro-Wilk test; $p<0.05$). Through Dunn's test, it is possible to confirm that there are no significant differences between AM and WM ($p>0.05$). However, there are significant differences among the rest of the evaluation approaches ($p<0.05$). Thus, a performance rank of the evaluation approaches can be performed: WM and AM $>$ GM $>$ HM.

When $n=5$ (see Figure~\ref{fig:experimental_coop_conf_5}), all the approaches exhibit a steady evolutionary behaviour. Nevertheless, their performance tends to evolve similarly. Moreover, AM, GM, and HM present scarce fluctuations during evolution. Significant differences in the performance of the evaluation approaches were confirmed. Normality testing using the Shapiro–Wilk test showed that the data were not normally distributed ($p < 0.05$). By utilising Dunn's test, it can be confirmed that no significant differences among WM, GM, and HM exist ($p>0.05$). On the other hand, there are significant differences between AM and the rest of the evaluation approaches ($p<0.05$). Therefore, ranking the performances is suitable: AM $>$ WM and GM and HM. 

In addition, when $n=7$ (see Fig.~\ref{fig:experimental_coop_conf_7}), again, a steady evolutionary behaviour is noticeable. Furthermore, WM tends to evolve faster than the rest of the approaches; however, in the last generations, AM, GM, and HM reach the same pace as WM. Furthermore, fluctuations are observed in the performance of AM, GM, and HM during evolution. Significant differences in performance were identified. Normality testing using the Shapiro–Wilk test indicated that the collected data were not normally distributed ($p < 0.05$). Employing Dunn's test makes it feasible to confirm no significant differences among AM, GM, and HM ($p>0.05$). In contrast, there are significant differences between WM and the rest of the approaches. These results enable a comparative ranking of the performance across the four evaluation approaches: WM $>$ AM and GM and HM.

When $n=10$ (see Fig.~\ref{fig:experimental_coop_conf_10}), although all the evaluation approaches exhibit a clear evolutionary process, they present several fluctuations. Moreover, statistical analysis revealed significant differences in performance among the four evaluation approaches. Thus, a performance rank can be performed: WM $>$ HM $>$ AM $>$ GM (Shapiro-Wilk test; Dunn's test: $p<0.05$).

\begin{figure}[tb!]
  \centering
     \includegraphics[width=1.0\linewidth]{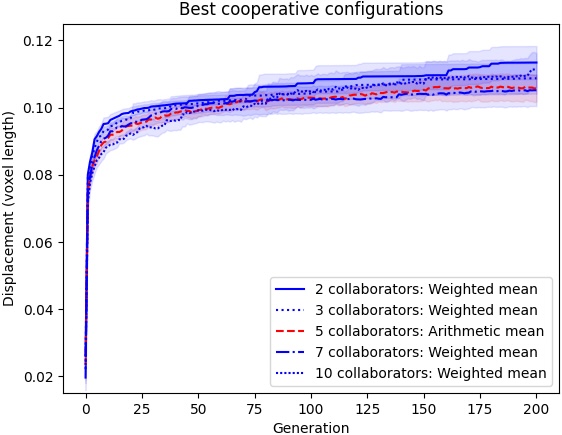}
  \caption{Mean performance observed of the fittest SAM showing 95\% confidence interval (shaded region) under the best cooperative configurations found.}
  \label{fig:experimental_coop_conf_best}
\end{figure}

For research completeness, Fig.~\ref{fig:experimental_coop_conf_best} exhibits the mean performance of the fittest individual with 95\% confidence intervals presented by the shaded regions under the best cooperative configurations. As expected, there are some significant differences among the performances of the cooperative configurations. Dunn's test confirmed that there were no statistically significant differences ($p > 0.05$) between: (i) WM (3 collaborators) and WM (10 collaborators); (ii) AM (5 collaborators) and WM (7 collaborators). In contrast, there are significant differences between WM (2 collaborators) and the rest of the cooperative configurations ($p<0.05$). Thus, it is possible to rank the performances: WM (2 collaborators) $>$ WM (3 collaborators) and WM (10 collaborators) $>$ AM (5 collaborators) and WM (7 collaborators).

These results suggest that when WM is implemented, it induces a suitable performance in finding SAMs (and their controllers) capable of maximising the displacement in the $yz$ plane. Furthermore, under WM, when a few individuals ($n=2$) or a significant number of individuals ($n=10$) are used in one population to evaluate all individuals in the other population, it positively impacts the evolutionary process. On the other hand, when the number of individuals used to evaluate the individuals of the other population is neither scarce nor excessive, the evolutionary process exhibits poor performance. This behaviour is arguably due to the weights used during experimentation (see Table~\ref{tab:setup_coevolutionary_weighted}), which highly rewards a substantial displacement when $n$ is in the $\{2,3,10\}$ set, inducing a faster evolution. In contrast, a significant displacement is moderately rewarded when $n$ is in the $\{5,7\}$ set, provoking a slower evolution. 

%=====
\subsection{Multi-objective vs NeuroEvolution vs NeuroCoEvolution}\label{sec:experimental_robustness}

SAMs represent devices that will be used in real environments (e.g., blood vessels); consequently, the control signals towards SAMs may be interfered with or distorted by the intrinsic noise of these scenarios. This noise can be seen as an element that induces a constant change in the state of the environment. Thus, the objective of this experiment is to assess the robustness, in terms of controllers being applied, of the fittest morphology found in the previous experiment (see Section \ref{sec:experimental_coop_conf}). 

To provide a baseline, two morphologies are included during experimentation: (i) the first morphology was found by Age-Fitness Pareto Optimisation (AFPO) \cite{Schmidt2010}, a multi-objective algorithm that aims to prevent premature convergence in numerous evolutionary approaches, and (ii) the second morphology was found by CPPN-NEAT under a single population framework (i.e., standard NeuroEvolution). Both morphologies, which are the fittest found, were taken from \cite{Alcaraz2024actuator}.

\begin{figure}[tb!]
  \centering
     \includegraphics[width=1.0\linewidth]{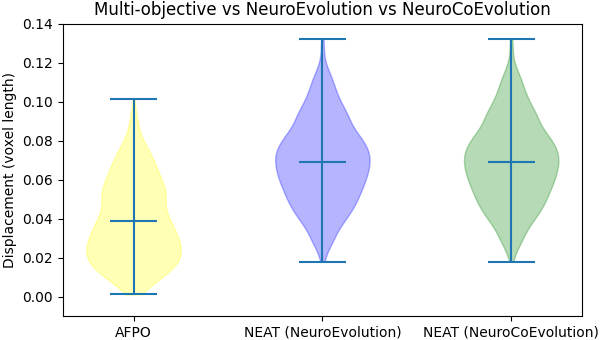}
  \caption{Displacement observed in the $yz$ plane under 1000 phase offset scenarios of the fittest SAM found by AFPO (left); CPPN-NEAT under NeuroEvolution (centre); and CPPN-NEAT under NeuroCoEvolution (right).}
  \label{fig:experimental_robustness}
\end{figure}

All SAMs are tested using 1000 different controller phase offsets, which were produced {\em a priori}. Figure~\ref{fig:experimental_robustness} depicts violin plots that compare the displacement observed in the $yz$ plane of the three SAMs found by AFPO, CPPN-NEAT under a NeuroEvolution approach, and CPPN-NEAT under a NeuroCoEvolution approach. Each violin plot illustrates the median, minimum, and maximum values, along with the kernel density estimation of value frequencies across 1,000 distinct controller phase offset scenarios.

Significant differences in displacements of the three SAMs were observed. Normality testing using the Shapiro–Wilk test indicated that the data were not normally distributed ($p < 0.05$). Subsequently, Dunn's test revealed no statistically significant differences in displacement between the morphologies produced by CPPN-NEAT using the NeuroEvolution-based approach and those generated using the NeuroCoEvolution-based approach ($p > 0.05$). In contrast, significant differences exist between the displacements produced by these morphologies and the AFPO-based morphology ($p<0.05$).

Results indicate that NeuroEvolution and NeuroCoEvolution can discover more robust and suitable SAMs than AFPO (i.e., multi-objective optimisation). Arguably, the capabilities of CPPN-NEAT to generate movement in soft materials \cite{Cheney2014} allow NeuroEvolution and NeuroCoEvolution to produce more bendable SAMs than AFPO. Moreover, despite the absence of significant differences between the two advanced methods in terms of producing SAMs that maximise the displacement in the $yz$ plane, NeuroCoEvolution was able to reach the same results in less time and fewer individuals. Whereas NeuroEvolution required 3000 generations with 100 individuals (and corresponding Voxelyze simulation runs) to produce the optimised SAM \cite{Alcaraz2024actuator}, NeuroCoEvolution took 200 generations with 50 individuals (25 SAMs and 25 controllers and corresponding Voxelyze simulation runs for 2 collaborators per individual) to generate a similar SAM (see Section~\ref{sec:setup_coevolutionary_setup}). To clearly illustrate the disparity in wall-clock time between the two approaches, NeuroCoEvolution completed the execution of one run in approximately 7.5 minutes, whereas the NeuroEvolution required nearly 20.5 hours. This stark contrast highlights the significant difference in computational efficiency and scalability between the two techniques.

%%%%%%% TO DO %%%%%%%%%%%%%%%%%
%Evolution took 300000 evaluations and co-evolution took 10000 evaluation. We can add rougly mention the wall clock time required per experiment. 
%Coevolution (1 run; WM with 2 collaborators): 456.72 sec, 7.61 minutes, 0.1268 hours.
%Evolution (1 run): 73,725.50 sec, 1,228.75 minutes, 20.48 hours.  
%%%%%%%%%%%%%%%%%%%%%%%%%%%%%%%%%%

\begin{figure}[tb!]
  \centering
     \includegraphics[width=1.0\linewidth]{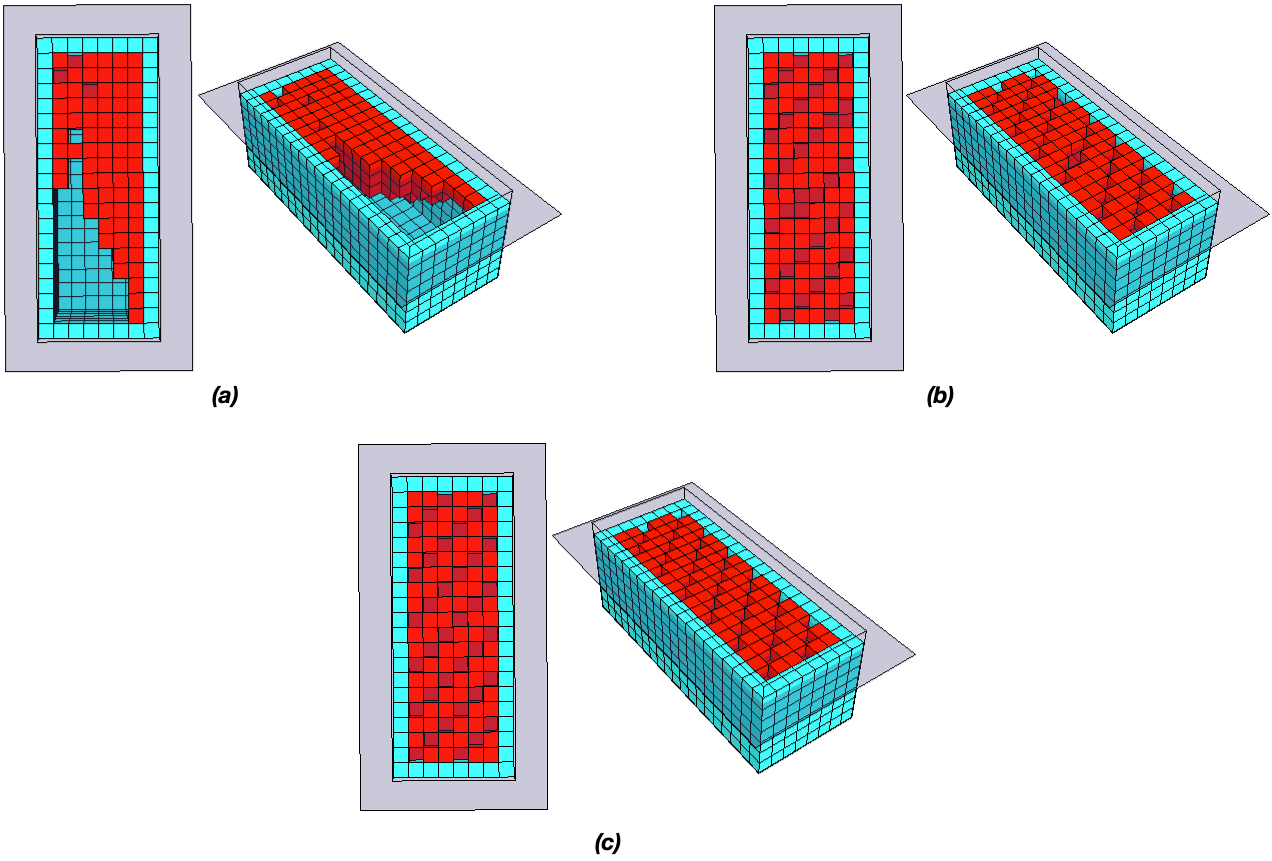}
  \caption{Fittest SAM found by: (a) AFPO; (b) NeuroEvolution; and (c) NeuroCoEvolution. Contractile tissue is represented by red voxels, whereas, passive tissue is indicated by blue voxels.}
  \label{fig:experimental_robustness_morphologies}
\end{figure}

Finally, Fig.~\ref{fig:experimental_robustness_morphologies} depicts the SAMs used in this experiment. A pyramidal-like pattern can be observed in the SAM discovered by AFPO (see Figure~\ref{fig:experimental_robustness_morphologies}-a). There is a gradual increase in voxel density from the base to the top of the SAM. Notably, the absence of voxels near the base appears to significantly diminish its bending capabilities. In addition, the SAM found through NeuroEvolution (see Figure~\ref{fig:experimental_robustness_morphologies}-b) shows a solid striped diagonal pattern of voxels. Appreciating two thick stripes in the middle of the SAM is possible.
Furthermore, the SAM found by NeuroCoEvolution (see Figure~\ref{fig:experimental_robustness_morphologies}-c) presents a similar voxel pattern observed in the SAM found by NeuroEvolution. However, the voxel stripes are thin throughout the SAM. Possibly, the diagonal voxel pattern exhibited by these SAMs allows them to produce suitable bending movements.

%%%%%%%%%%%%%%%
% CONCLUSIONS %
%%%%%%%%%%%%%%%

%=======
\section{Conclusions}\label{sec:conclusions}

This research analysed CPPN-NEAT as a design engine for SAMs and controllers under a cooperative CoEvolutionary scheme. Two experiments were conducted to perform the analysis: (a) determining the optimal cooperative configuration able to generate suitable SAMs (and controllers) capable of maximising the displacement in the $yz$ plane, and (b) investigating the robustness of the previously identified optimal SAMs.

For the first experiment, four evaluation strategies to compute the fitness value of individuals were used based on: (i) the arithmetic mean; (ii) the weighted mean; (iii) the geometric mean; and (iv) the harmonic mean. One collaboration strategy was used: {\em n fittest individuals vs all}. The value of $n$ is assigned to each value of the $\{2, 3, 5, 7, 10\}$ set. For each cooperative configuration, 30 evolutionary runs were performed. 

Regarding the second experiment, the fittest SAM found in the first experiment was compared against two SAMs discovered in previous research. The first SAM was found by AFPO, a well-known multi-objective optimisation algorithm, whereas CPPN-NEAT found the second SAM using a NeuroEvolution-based approach. The benchmark focused on comparing the robustness of the three SAMs employing 1000 different phase offset scenarios. 

Results point out that regardless of evolutionary scheme (i.e., evolution or CoEvolution), CPPN-NEAT can surpass AFPO in terms of finding suitable SAMs due to its core mechanism being based on CPPNs, which allow to generate morphologies that exhibit patterns such as the ones observed in  Figure~\ref{fig:experimental_robustness_morphologies}-b and Figure~\ref{fig:experimental_robustness_morphologies}-c. These morphological patterns enhance the bending movement of SAMs. 

Furthermore, under a cooperative NeuroCoEvolution scheme, the weighted mean was the most suitable approach (in particular, when $n=2$) to evaluate individuals because its mechanism rewards a significant displacement. In contrast, a minimal displacement is severely ``punished''. Although this cooperative approach showed a suitable performance, it could not outperform but matched the performance of the NeuroEvolution approach. There is, nevertheless, one insight that is worth highlighting: NeuroCoEvolution was capable of obtaining the same performance (i.e., an evolved SAM capable of maximising bending movements in the $yz$ plane) as NeuroEvolution with fewer generations and fewer individuals, thus less computational resources (see Section~\ref{sec:experimental_robustness}). This behaviour can be understood as NeuroCoEvolution being more efficient in exploring the search space because a symmetrical ``evolutionary pressure'' is generated between SAMs and controllers. 

Based on the insights and findings obtained in this study, several directions for future research can be identified. For instance, more physical parameters such as viscosity and friction should be included during simulations to provide more realistic scenarios. Moreover, under the same cooperative CoEvolutionary scheme, the implementation of other NeuroEvolution-based algorithms such as Hypercube-based Neuroevolution of Augmenting Topologies (HyperNEAT) \cite{Stanley2009} and ES-HyperNEAT \cite{Risi2012}, which extends from HyperNEAT, is another potential direction for future work.

%Furthermore, other NeuroEvolution-based approaches can be used to conduct a more robust benchmark. For instance, Evolutionary eXploration of Augmenting Memory Models (EXAMM) aims to evolve ANNs by employing numerous memory structures \cite{Ororbia2019,Thakur2023}.

Finally, a relevant insight is that the NeuroCoEvolution system described in this research has three main components: SAMs, controllers and the area where SAMs are simulated. These three elements can be understood as {\em body}, {\em brain}, and {\em environment}, which are key concepts of the computational approach known as {\em embodied intelligence} \cite{Cangelosi2015}. Thus, this research can also be seen as an early step towards this research area.

\section*{Acknowledgement}
This project has received funding from the European Union’s Horizon Europe research and innovation programme under grant agreement No. 101070328. UWE researchers were funded by the UK Research and Innovation grant No. 10044516.

\end{document}